%% file: tempelet.tex
\documentclass{article}
\usepackage{spconf,amsmath,graphicx}

\usepackage{xcolor}
\usepackage{multirow} 
\usepackage{booktabs}
\usepackage{amssymb}

\title{Exploring Multi-Modal Control in Music-Driven Dance Generation}
\name{{Ronghui Li}$^{1,2 \dagger}$, {Yuqin Dai}$^{3 \dagger}$, {Yachao Zhang}$^{1 \star}$, {Jun Li}$^{3}$,{Jian Yang}$^{3}$, {Jie Guo}$^{2}$, {Xiu Li}$^{1 \star}$
\thanks{${\dagger}$ These authors contributed equally to this work.}  
\thanks{${\star}$ Corresponding authors.}
\thanks{$\textcopyright$ 2024 IEEE.  Personal use of this material is permitted.  Permission from IEEE must be obtained for all other uses, in any current or future media, including reprinting$/$republishing this material for advertising or promotional purposes, creating new collective works, for resale or redistribution to servers or lists, or reuse of any copyrighted component of this work in other works.}
}
\address{$^1$Shenzhen International Graduate School, Tsinghua University, 
$^2$Peng Cheng Laboratory, \\
$^3$School of Computer Science and Engineering, Nanjing University of Science and Technology
}

\begin{document}
%
\maketitle
\begin{abstract}
\input{sections/Abstract}
\end{abstract}
\section{Introduction}
\label{sec:intro}
\input{sections/Introduction}

\section{Method}
\label{sec:Method}
\input{sections/Method}
\section{Experiment}
\label{sec:Experiment}
\input{sections/Experiment}
\section{Conclusion}
\label{sec:Conclusion}

In this paper, we propose a unified framework capable of generating high-quality dance and supporting the control of genre, text, and keyframe. 
We solve the issue of quality degradation caused by the introduction of control information. 
Experiment results demonstrate that our method outperforms existing networks in both dance quality and controllability.

\section*{Acknowledgements}
This research was partly supported by Shenzhen Key Laboratory of next generation interactive media innovative technology (Grant No: ZDSYS20210623092001004), the China Postdoctoral Science Foundation (No.2023M731957), the Peng Cheng Laboratory (PCL2023A10-2), the National Natural Science Foundation of China under Grant 62306165, 62072242 and 62361166670.

\vfill\pagebreak
\bibliographystyle{IEEEbib}
\bibliography{ref}

\end{document}

%% file: sections/Abstract.tex
Existing music-driven 3D dance generation methods mainly concentrate on high-quality dance generation, but lack sufficient control during the generation process.
To address these issues, we propose a unified framework capable of generating high-quality dance movements and supporting multi-modal control, including genre control, semantic control, and spatial control. First, we decouple the dance generation network from the dance control network, thereby avoiding the degradation in dance quality when adding additional control information. Second, we design specific control strategies for different control information and integrate them into a unified framework. 
Experimental results show that the proposed dance generation framework outperforms state-of-the-art methods in terms of motion quality and controllability.


%% file: sections/Introduction.tex

In today's era of digital entertainment, 
there is a growing need for the efficient generation of high-quality, controllable 3D dances based on provided music. 
With the development of AIGC technology ~\cite{ma2023follow,lin2023consistent123}, this is becoming a reality.
However, most existing works focus on the dance quality while neglect the controllability.  

Early methods \cite{aist++,dancerevolution}
input music and seed motions into a single network, such as Transformer \cite{attention}, generating new dance movements frame by frame in an autoregressive manner. However, challenges of error accumulation and motion freezing phenomena still persist.
Recently, some methods can generate high-quality dance based on music.
Bailando \cite{bailando} trains a VQ-VAE network to encode dance motion segments into tokens. Subsequently, a Transformer is used to predict dance token sequences from input music, ultimately decoded into 3D dance by a VQ-VAE Decoder. Additionally, Bailando introduced an Actor-Critic network to enhance the quality of dance actions and mitigate the motion-freezing issues present in the previous methods.
FineDance \cite{FineDance} and EDGE \cite{edge} utilize the diffusion \cite{ho2020denoising} model for dance generation, resulting in high-quality and diverse dance sequences. 

\label{sec:gpt}
\begin{figure}[tbp]
  \centering
  \mbox{} \hfill
  \includegraphics[width=1.0\linewidth]{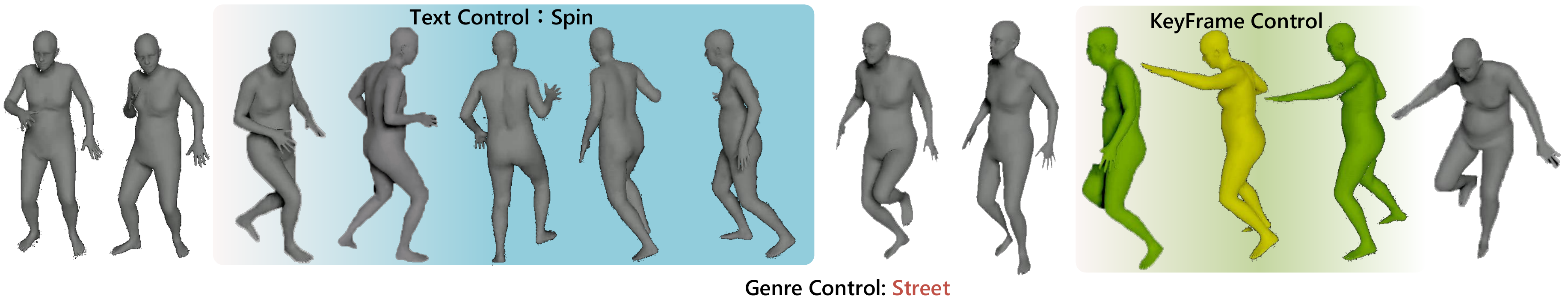}
  \vspace{-8mm}
  \caption{{Generated dance from various control input and music. 
  }
  \vspace{-3mm}
  \label{fig:fullcontrol}%
   }
\vspace{-5mm}
\end{figure}

However, existing methods have not sufficiently explored the controllability of dance generation. 
In practical dance composition, choreographers have the ability to control the genre, semantic, and spatial details of the dance.
Different control signals such as genre control \cite{mnet}, text-based semantic control \cite{tm2d}  and keyframe-based spatial control \cite{edge}, \textit{there is still a lack of a unified framework to control the genre, semantics, and spatial details of dance simultaneously.}


\begin{figure*}[tbp]
  \centering
  \mbox{} \hfill
  \includegraphics[width=1.0\linewidth]{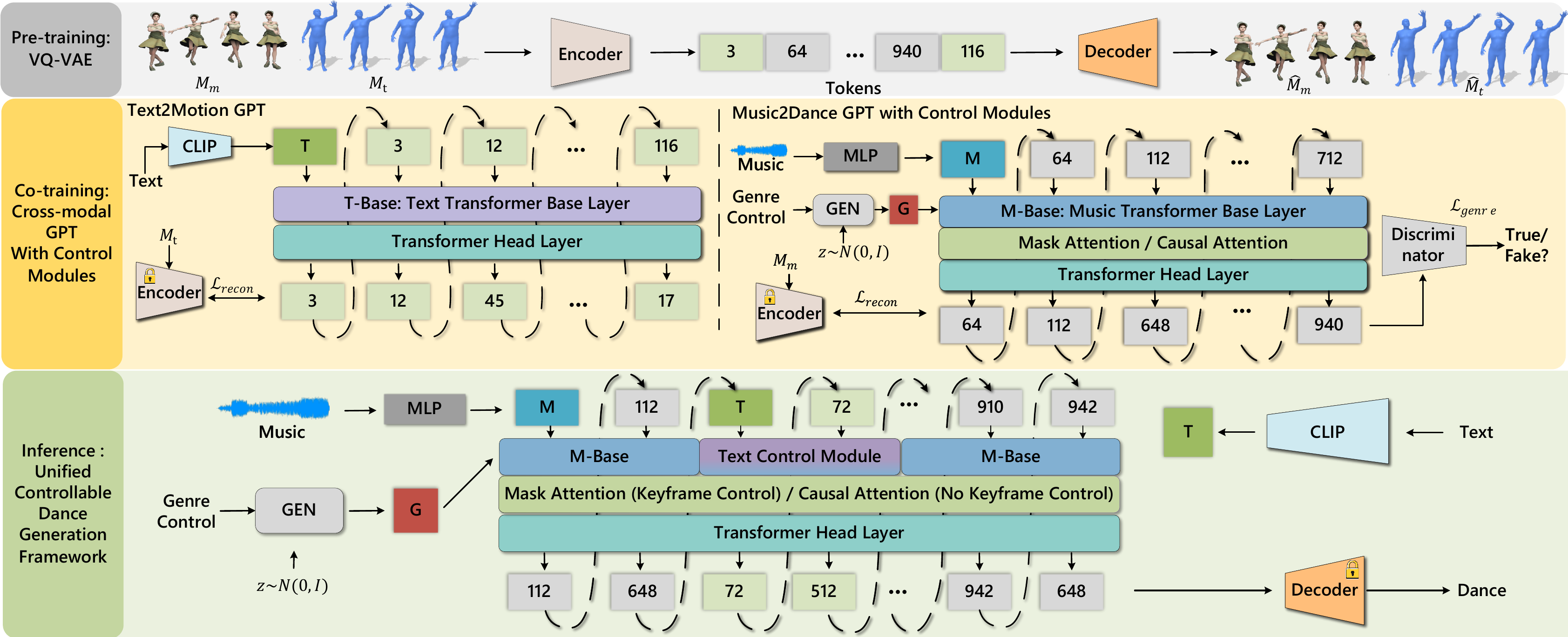}
  \vspace{-8mm}
  \caption{Overview of our method. GEN means Genre Embedding Network.
  \label{fig:framework}%
   }
\vspace{-4mm}
\end{figure*}

Generating multi-modal controllable dance faces two key challenges: (1) How to ensure both effective control and high-quality dance generation? In previous approaches, dance control and dance generation are tightly coupled, which results in a degradation of dance quality when control signals are introduced. This issue becomes more prominent when multiple modalities of control are concurrently integrated.
(2) How to achieve multi-modal control within a unified framework? 
The genres, text, and keyframes represent three entirely distinct modalities, posing significant challenges to the network's modeling capabilities due to the huge modal gap and the abundance of input signals.

To solve the above issues,
We decouple the dance generation from dance control by pretraining a VQ-VAE. Its encoder can transform dance clips into tokens, 
and then are used to reconstruct dances.
We constrain the control network to predict only these tokens, effectively fixing the VQ-VAE parameters to ensure the quality of generated dance movements. To achieve effective control guided by multi-modal input signals, we design a controllable dance token prediction network based on the GPT architecture \cite{attention, gpt}. We integrate multi-genre embedding network and multi-genre discriminators to achieve genre control. We also design a shared latent space for text and music and fused their features for semantic control. 
Additionally, we utilize GPT's mask prediction strategy for keyframe control. Finally, our method offers flexibility in achieving controllable dance generation for one or multiple modalities and demonstrates promising results in both qualitative and quantitative experiments.

Our main contributions can be summarized as follows: (1) We propose a unified framework that can generate dance from given music while supporting genre, text, and keyframe control.
(2) We decouple the dance generation network from the dance control network, achieving both control effectiveness and the generation of high-quality dances.

%% file: sections/Method.tex
\textbf{Problem Definition.}
Given the input music, our goal is to generate high-quality dance while allowing for control over genre, text, and keyframes simultaneously. 
Given a genre $g$, a text prompt $p$, a piece of motion keyframes $m_k$, and the music feature $y\epsilon\mathbb{R}^{N\times {C_m}}$ extracted by librosa \cite{librosa}, where $N$ is the feature length and $C_m$ is the feature dimension.
Our method can generate dance
$x\epsilon\mathbb{R}^{N\times {C_d}}$ ($C_d$ is the dance feature dimension,
corresponding to $y$, while obey the control of $g, p, m_k$. 
The overview of our framework is shown in Fig.~\ref{fig:framework}. 

\noindent\textbf{Method Overview.}
First, we train a VQ-VAE capable of projecting motion to tokens and vice-versa. Second, we train the basic music2dance GPT and text2motion GPT on the paired tokens and music/text feature. During this period, we alternately train two GPT models and share weight of Transformer head layer for the preparation of text control.
Then, we train the genre control network of the Genre Embedding Network and Multi-genre Discriminator and the keyframe control module of MaskAttention layer. Finally, we can use the unified framework to generate dance under multi-control.

\subsection{Pre-training: Motion VQ-VAE}
\label{sec:vqvae}
There is no existing text\&music2dance dataset, but we have $M_m=\left\{x|x\, \text{is music-paired dance}\right\}$ and we also have  $M_t=\left\{x|x\,\text{is text-paired motion}\right\}$. 
We utilize a VQ-VAE to project $M_m \cup M_t$ into a codebook, which is a shared latent space for $M_m$ and $M_t$. In this way, all the motions can be transformed into tokens.
To prevent the degradation in dance quality caused by the addition of control, we decouple dance generation and dance control. We fix the parameters of the trained VQ-VAE and only use the token sequences obtained from the VQ-VAE's encoder to train the dance control network.

\subsection{Training basic Cross-modal GPT}
We employ Cross-modal GPT as a basic model to achieve the following task: 
\textbf{Text to motion:} 
For training text2motion GPT, we follow the \cite{t2mgpt} to maximize the log-likelihood of the data distribution:
\vspace{-1.5mm}
\begin{equation} 
\mathcal{L}_{\text {recon }}=\mathbb{E}_{{x} \sim P({M_t})}[-\log p({x} \mid T)]
\vspace{-1.5mm}
\end{equation}
where $T$ is the text embedding extract by CLIP \cite{clip}. 

\noindent\textbf{Music to dance:} We use MLP  to extract music embedding $M$. The music2dance GPT is trained $\left\{Y, M_m\right\}$, where $Y=\left\{y|y\, \text{is dance-paired music}\right\}$. The training process is similar to text2motion GPT.
As Fig.\ref{fig:framework} shows, the GPT model consists of Transformer base and head layers, which are composed of linear and attention layers.
To prevent confusion with music and text features, we design two distinct base layers to extract the music/text features respectively. 




\subsection{Multi modal control}

\begin{figure}[tbp]
  \centering
  \mbox{} \hfill
  \includegraphics[width=1.0\linewidth]{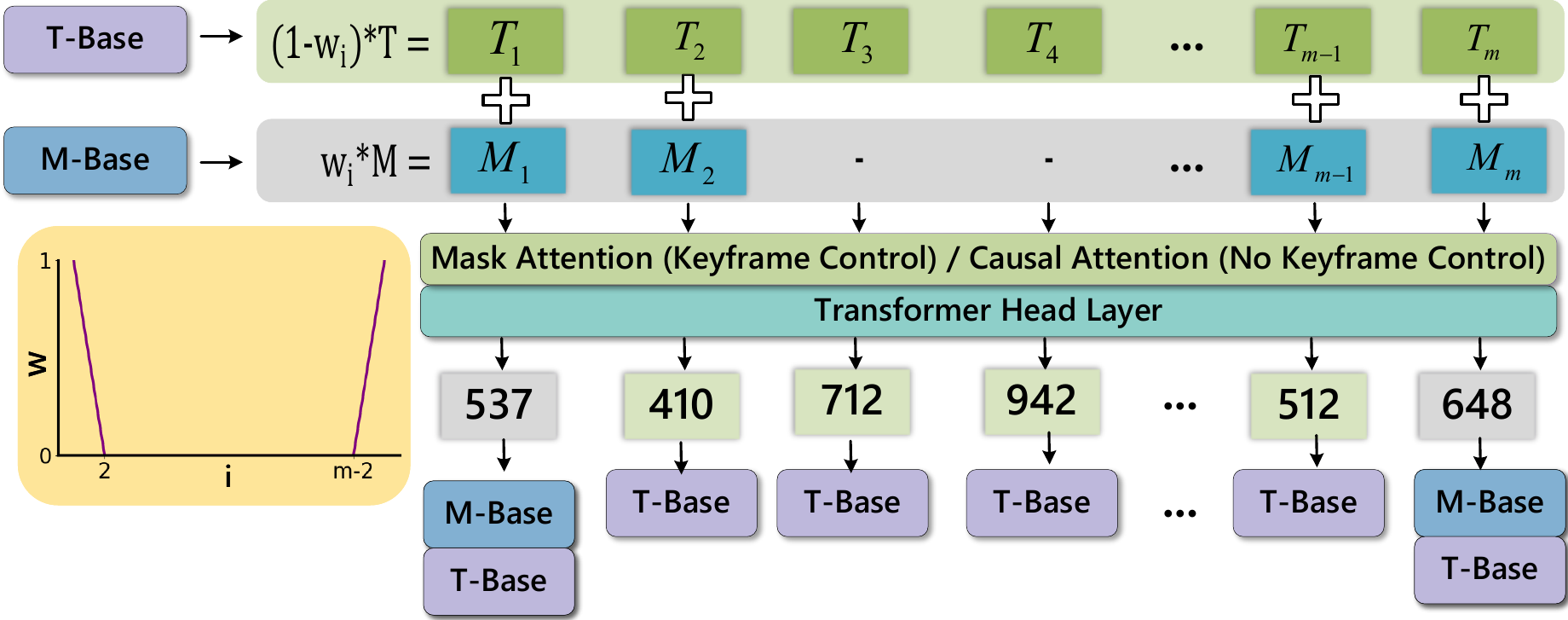}
  \vspace{-8mm}
  \caption{{Text control module.}
  \vspace{-6mm}
  \label{fig:fusion}%
   }
\vspace{-3mm}
\end{figure}

\textbf{Text Control.}
\label{sec:fusion}
Thanks to the VQ-VAE trained on $M_m \cup M_t$, capable of decoding semantically meaningful motions; and the text2motion GPT model trained on $\left\{M_t, p\right\}$, excels at extracting text features and predicting motion tokens.
We are able to introduce semantic control into our dance generation process.  
To make the transformer head layer generate tokens that possess semantic meaning and adhere to dance standards,
we alternately train the text2motion GPT and music2dance GPT while sharing the head layer to process features provided by the base layers and predict motion tokens.
However, directly incorporating semantic movements into the dance can significantly deteriorate the quality of the dance, resulting in severe incoherence. 
Therefore, we set a transitional interval in which we fusion music features $M$ extracted from the M-base and text features $T$ extracted from the T-base:
\vspace{-1.5mm}
\begin{equation}
F_i=T_i*(1-w_i)+M_i*w_i
\vspace{-1.5mm}
\end{equation}
where $F_i$ is the fusion feature and $w_i$ is the fusion weight. The weight change pattern provided by Fig.\ref{fig:fusion}.
The fused feature $F_i$ is inputted to the head layer to predict the dance tokens with the guided text. Finally, using the pre-trained VQ-VAE decoder can reconstruct the 3D dance from tokens.

\label{sec:gpt}

\noindent\textbf{Genre Control.}
We use M-Base layer for cross-modal music feature extraction and implementing genre control during the feature extraction stage. 
We employ a genre embedding network \cite{mnet} to embedding the genre $g$ and random code $\mathbf{z}$ into $G$ and use a cross-attention layer to model the features:
\vspace{-1.8mm}
\begin{equation}
\operatorname{CrossAttention}=\operatorname{Softmax}\left(M G^T / \sqrt{d}+B\right) G
\vspace{-1.8mm}
\end{equation}
where $B$ is the bias, and $d$ is a scaling factor to ensure the stability of the model's training process.
We use MLP for the extraction of music feature $M$ and feed the music features into GPT sequentially. 
The training process of music2dance with genre control can be formulated as:
\vspace{-1.5mm}
\begin{equation}
\begin{aligned}
\mathcal{L}_{\text {genre }}= & \mathbb{E}_{x \sim P({M_m})}\left[\log D\left(x, g, y\right)\right]+ \\
& \mathbb{E}_{\mathbf{z} \sim \mathcal{N}(0, \mathbf{I})}\left[\log \left(1-D\left(GPT\left(\mathbf{z}, g, y\right), g, y\right)\right)\right],
\end{aligned}
\vspace{-1.5mm}
\end{equation}
where the $D(\cdot)$ is the multi-genre dance discriminator.

\noindent\textbf{Keyframe Control.}
\label{sec:Frame control}
Based on the GPT framework, we employ the mask and predict mechanism to achieve keyframe control. 
In the previous training processes, we generate token sequences step-by-step using causal attention. To achieve keyframe control, we replace causal attention with mask attention and train it to predict tokens that are randomly masked.
In inference, we first use the GPT model (with causal attention layer) to generate dance tokens based on the music. Subsequently, we encode the keyframe into tokens and insert them into the previously generated token sequence. We mask out the tokens of the keyframe to enable GPT  (with mask attention layer) to predict the before and after k motion tokens, thereby achieving keyframe control while ensuring the coherence of the generated dance sequence. 

\subsection{Inference: Unified Framework}
Each Control Module has been designed to be plug-and-play, allowing for the inclusion or removal of each control signal as desired. Once the training is complete, it becomes convenient and flexible to control the dance sequences we wish to generate. 
By modifying the attention layer of casual attention or mask attention, different functions such as sequence generation and keyframe control can be achieved.


%% file: sections/Experiment.tex
\begin{figure}[tbp]
  \centering
  \mbox{} \hfill
  \includegraphics[width=1.0\linewidth]{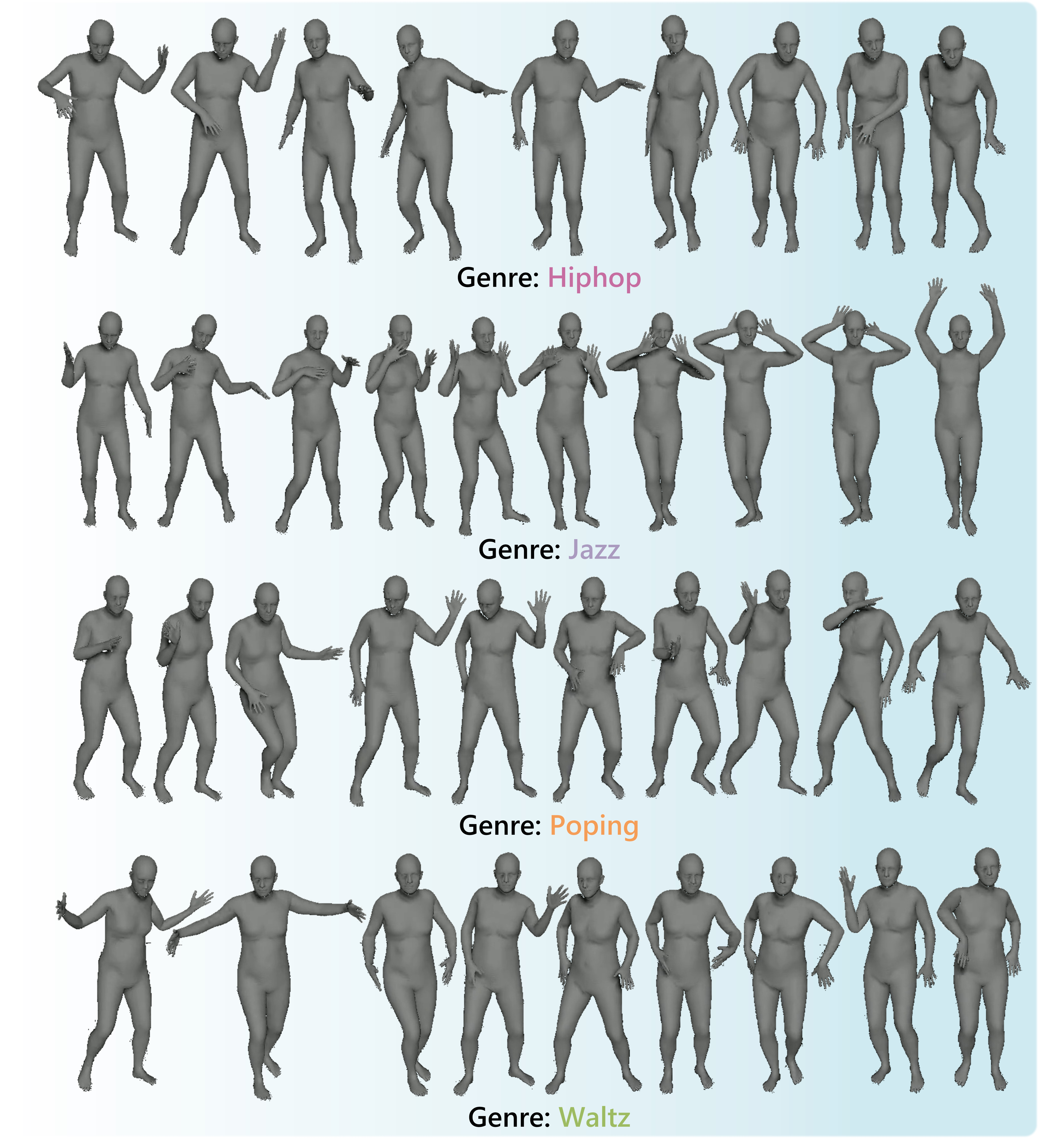}
  \vspace{-8mm}
  \caption{Generated dance for the same music in different genres, showcasing the effective control of the given genre on the generated sequence and the diversity achieved.
  \vspace{-3mm}
  \label{fig:genre}%
   }
\vspace{-1.5mm}
\end{figure}

\subsection{Setups}
\textbf{Data Processing.} 
Finedance \cite{FineDance} is a music2dance dataset with 22 fine-grained genres. HumanML3D \cite{humanml3d} is a text2motion dataset. 
We preprocess to make them balance and unify the motion data format of SMPL \cite{loper2015smpl} with 22 joints.

\noindent\textbf{Implementation Details.} 
The codebook size of VQ-VAE is $1024 \times 512$.  For both HumanML3D \cite{humanml3d} and Finedance \cite{FineDance} datasets, the motion sequences are concatenated or cropped to $t = 128$ for training. 
The parameter $k$ is set to 6.


\begin{figure}[tbp]
  \centering
  \mbox{} \hfill
  \includegraphics[width=1.0\linewidth]{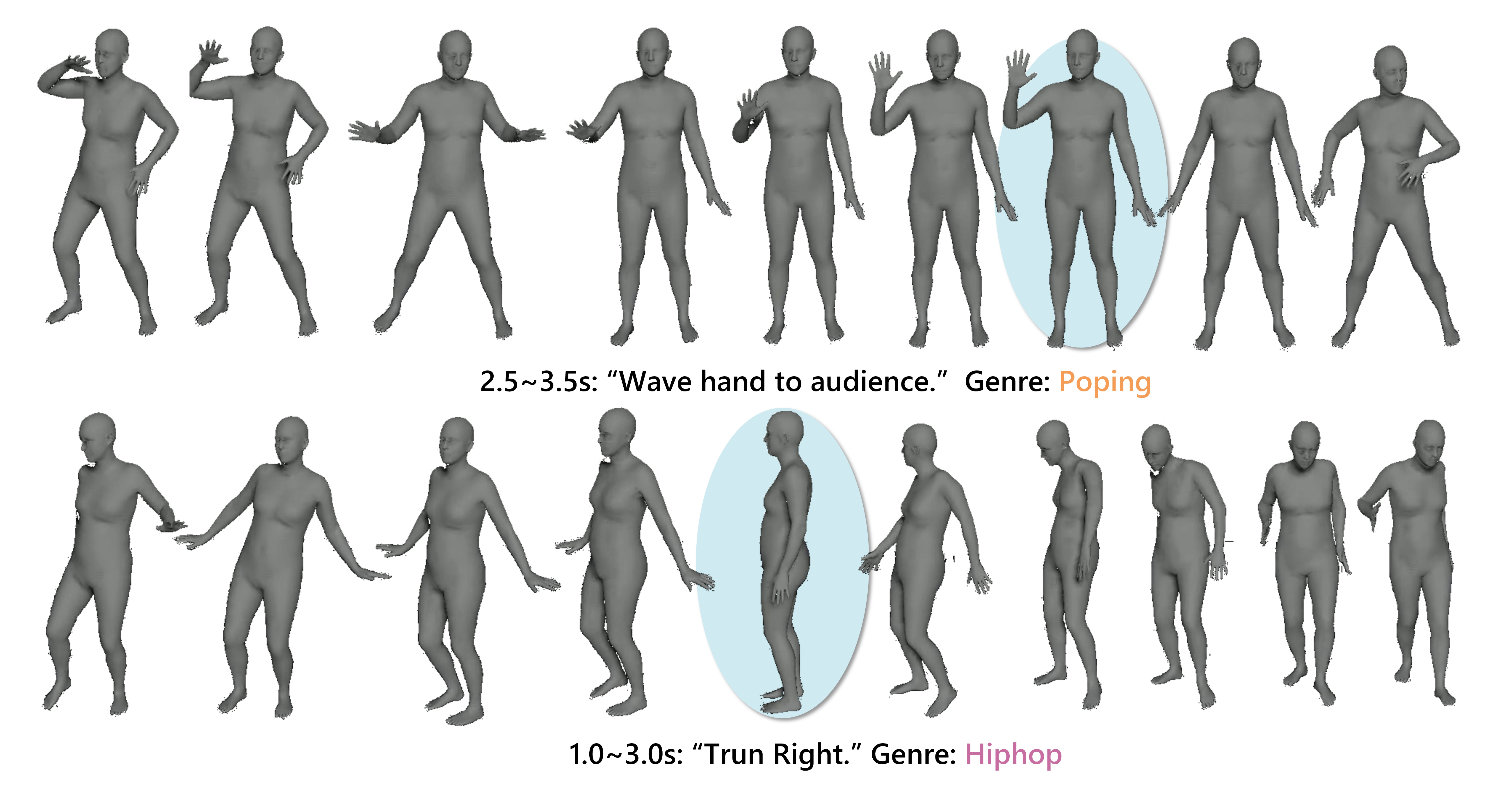}
  \vspace{-9mm}
  \caption{
  Generated dances for the same music using different text controls.
  \label{fig:motion}%
   }
\vspace{-3mm}
\end{figure}

\vspace{-1mm}
\subsection{Comparative Results} 
\begin{figure}[tb]
  \centering
  \mbox{} \hfill
  \includegraphics[width=1.0\linewidth]{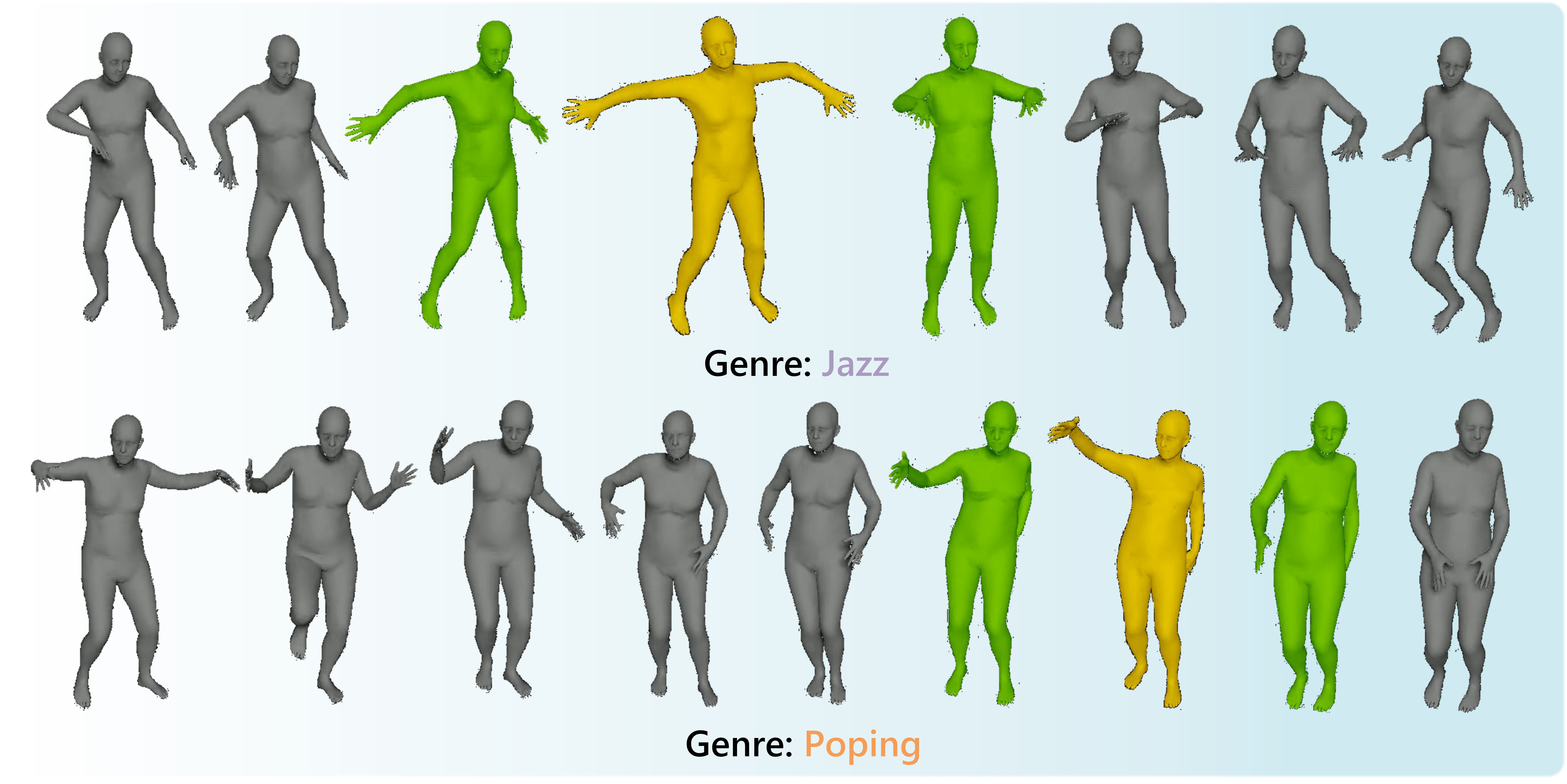}
  \vspace{-8mm}
  \caption{Keyframe control results of our model, yellow parts denote the keyframes and green parts denote the mask-predicted motion. The results demonstrate its exceptional capability to accurately predict a cohesive sequence of actions by taking into account the contextual information, thus effectively achieving keyframe control.
  \vspace{-3mm}
  \label{fig:keyFrame}%
   }
\end{figure}

\begin{figure}[tb]
  \centering
  \mbox{} \hfill
  \includegraphics[width=1.0\linewidth]{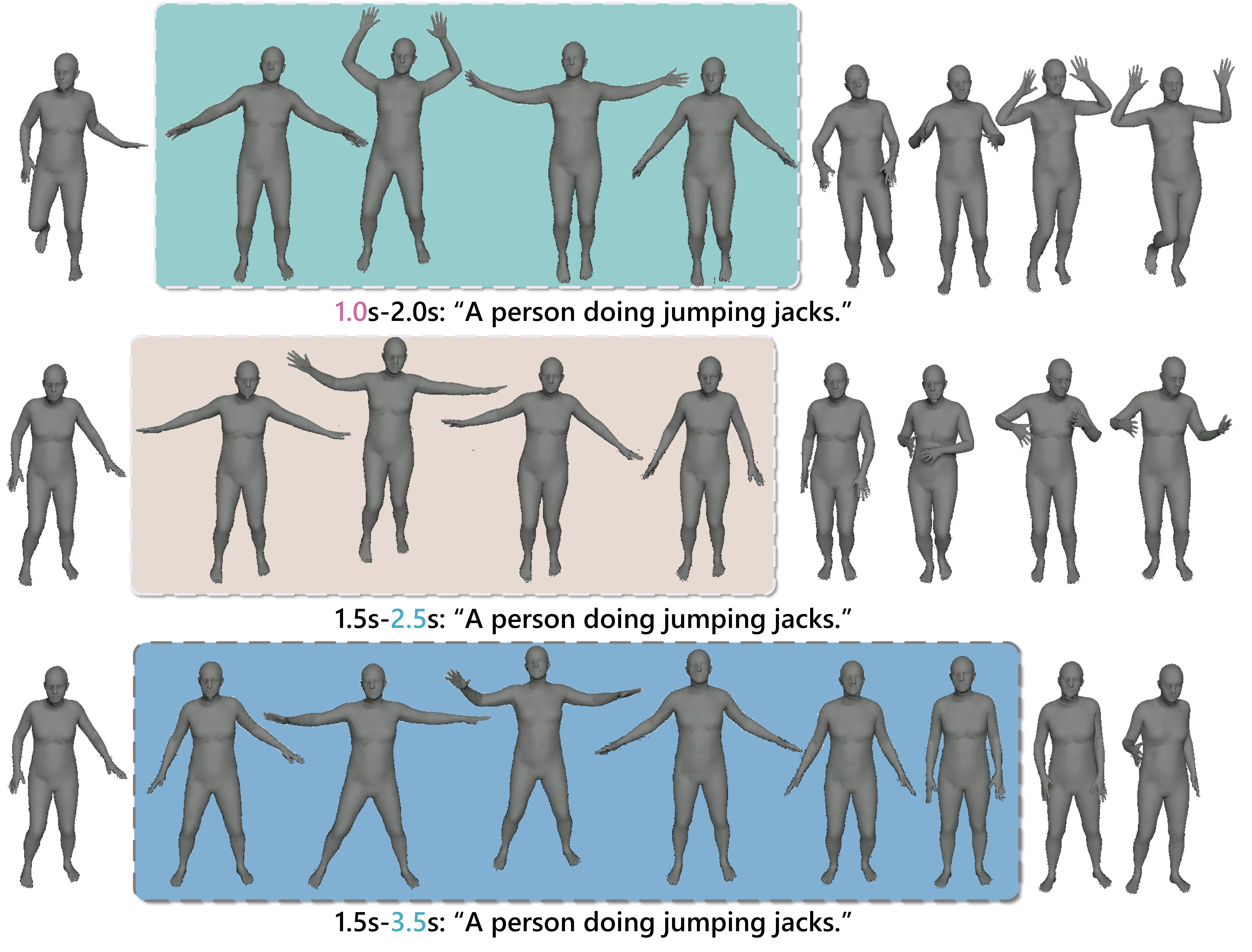}
  \vspace{-9mm}
  \caption{Generated dances of the same text control under different durations, further indicating that our model can effectively adapt to different action transitions.
  \label{fig:duration}%
   }
\end{figure}

\textbf{Qualitative Results.}
Fig.\ref{fig:fullcontrol} shows the combined control effects of genre, text, and keyframe inputs, while Fig.\ref{fig:genre}, \ref{fig:motion}, and \ref{fig:keyFrame} respectively demonstrate the control effects of genre, text, and keyframes. Fig.\ref{fig:duration} showcases the enhancement of dance diversity resulting from the introduction of different control signals.
We compare the user preference of our method with other SOTA methods. Each subject is asked to watch randomly presented videos and assign separate ratings from 1 to 5 for motion quality, fluency, and control effectiveness.

\noindent\textbf{Quantitative Comparisons.} 
We follow the settings of Bailando \cite{bailando} to evaluate the dance generation quality, including FID \cite{fid} and Diversity. 
The subscripts $k$ and $g$  represent kinetic feature \cite{onuma2008fmdistance} and geometric \cite{muller2005efficient} feature respectively.


\begin{table}[ht]
\vspace{-1mm}
\caption{Comparisons of motion quality and diversity.}
\label{tab:performance_m2d}
\centering
\setlength{\tabcolsep}{3.5pt} 
\begin{tabular}{lcccccc}
\toprule
    \multirow{2}{*}{\textbf{Methods}} & \multicolumn{2}{c}{\textbf{\underline{Motion Quality}}} & \multicolumn{2}{c}{\textbf{\underline{Motion Diversity}}}\\
    \noalign{\smallskip}
    & $FID_k \downarrow$ & $FID_g \downarrow$ & $Div_k \uparrow$ & $Div_g \uparrow$ \\
\midrule
Ground Truth & - & - & 10.03 & 7.37 \\
\hline
DanceRevolution \cite{dancerevolution} & 380.38 & 339.72 & 15.30 & 5.32 \\
DeepDance \cite{deepdance} & 256.77 & 177.42 & \textbf{31.71} & 1.95 \\
EDGE \cite{edge} & 51.90 & \textbf{40.49} & 9.20 & 9.04 \\
Ours & \textbf{38.56} & 53.08 & 7.41 & \textbf{9.32} \\
\bottomrule
\end{tabular}
\vspace{-1mm}
\end{table}

\begin{table}[!h]
\vspace{-1mm}
\caption{User study. KF means Keyframe control.}
\label{tab:user_study}
\centering
\setlength{\tabcolsep}{2.0pt} 
\begin{tabular}{lcccccc}
\toprule
     \multirow{2}{*}{\textbf{Methods}} & \multicolumn{3}{c}{\textbf{\underline{Accept Score}}} & \multicolumn{3}{c}{\textbf{\underline{Control}}}\\
     & Quality & Fluency & Ctrl & Text & Genre & KF\\
\midrule
DanceRevolution \cite{dancerevolution} & 3.10 & 3.40 & 2.80 &  & \checkmark &  \\
MNET \cite{mnet} & 3.90 & 3.70 & 4.00 &  & \checkmark & \\
Ours & 3.67 & \textbf{3.78} & 3.75 &  & \checkmark &  \\ %
\hline
TM2D \cite{tm2d} & 3.50 & 3.60 & 3.70 & \checkmark &  &  \\
Ours & \textbf{3.75} & 3.50 & \textbf{3.70} & \checkmark &  &  \\
\hline
EDGE \cite{edge} & 3.89 & 3.56 & 3.80 &  &  & \checkmark \\
Ours & 3.78 & \textbf{3.67} & 3.67 &  &  & \checkmark \\ 
\hline
Ours & 3.73 & 3.65 & 3.71 & \checkmark & \checkmark & \checkmark \\
\bottomrule
\end{tabular}
\vspace{-1mm}
\end{table}